# Opinions can be Incorrect! In our Opinion.
# On the accuracy principle in data protection law

Dara Hallinan[1] and Frederik Zuiderveen Borgesius[2]

1. Corresponding author: Dara Hallinan, IGR, FIZ Karlsruhe – Leibniz-Institut für Informationsinfrastruktur, Hermann-von-Helmholtz-Platz 1, 76344, Eggenstein-Leopoldshafen, Germany. E-mail: dara.hallinan@fiz-karlsruhe.de

2. Frederik Zuiderveen Borgesius, Institute for Computing and Information Sciences (iCIS), Radboud University, Nijmegen, Netherlands and Institute for Information law, University of Amsterdam, Amsterdam, Netherlands

- The GDPR contains an accuracy principle, as most data privacy laws in the world do. In principle, data controllers must ensure that personal data they use are accurate.

- Some have argued that the accuracy principle does not apply to personal data in the form of opinions about data subjects.

- We argue, however, from a positive law perspective, that the accuracy principle does apply to opinions.

- We further argue, from a normative perspective, that the accuracy principle should apply to opinions.

## I. Introduction

Since, the early 1970s, the accuracy principle has been a bulwark of data protection law. Roughly summarised, the principle says that organisations that use personal data must ensure that personal data are reasonably correct. The accuracy principle is included in many data protection and information privacy statutes around the world. The principle is also listed as one of the overarching data protection principles in the European General Data Protection Regulation (GDPR).[1]

---

[1] Regulation (EU) 2016/679 of the European Parliament and of the Council of 27 April 2016 on the protection of natural persons with regard to the processing of personal data and on the free movement of such data, and repealing Directive 95/46/EC (General Data Protection Regulation), OJ 2016 L 119/1, Article 5(1)(d).

Even despite the significance and track-record of the accuracy principle in data protection law, neither courts nor scholars have felt it necessary to spend much time or ink elaborating the content of the principle. Perhaps the principle is presumed to be simple and self-explanatory and thus without need for conscious exploration. As the UK Information Commissioner's Office has observed: 'It will usually be obvious whether personal data is accurate'.[2]

Yet, for some types of personal data, the applicability of the principle is far from self-explanatory. One such case is that of personal data in the form of opinions – also called inferences. The issue is not addressed in case law. Whilst the issue has been touched on by academic commentators, the depth of their analysis is limited and the conclusions they reach contradictory. This paper thus focuses on the following question: are opinions subject to data protection law's accuracy principle, and should they be?

Considering both pro and contra positions, we argue that opinions both are, and should be, subject to data protection law's accuracy principle. To keep the discussion in the paper manageable and concrete, we focus on one example of data protection law: the GDPR. Nevertheless, our analysis and conclusions are generally applicable to all privacy and data protection acts which include an accuracy principle – the majority of privacy and data protection laws the world over.

The focus of the paper is relevant for both practitioners and scholars for at least three reasons. First: the accuracy principle highlights the normative illegitimacy of the spread, wilful or negligent, of false or misleading information about individuals – it is the direct link between data protection and truth. Second: the processing of opinions occurs across a huge range of social contexts. Opinions appear whenever either humans or algorithms provide interpretations of existing facts or propositions, to derive new facts or propositions. Third: as noted, there has been little in-depth scholarly attention on the topic.

To start, the paper considers the accuracy principle in the GDPR. This consideration includes a look at the role and significance of the principle as well as a look at the content of the principle. Next, the paper discusses what an opinion is and highlights the huge range of contexts in which personal data in the form of opinions are processed. Finally, the paper presents arguments outlining that the accuracy principle does apply to opinions and that this applicability is a normatively desirable. We end with a conclusion.

## II. The Accuracy Principle in Data Protection Law

Since the 1970s, data protection law has become increasingly important, as personal data play an ever-larger role in our societies. Data protection law is, in large part, defined by a set of seven substantive data protection principles.[3] These overarching norms are called the 'principles relating to processing of personal data' in the GDPR. They are sometimes also called the Fair Information Principles – especially by US lawyers. Since their inception, they have shown themselves to be both durable as and scalable. Accordingly, they have proliferated rapidly around the world. Now, more than 120 countries have privacy laws with similar principles.[4] The accuracy principle is one of these seven key principles.

---

[2] See: <https://ico.org.uk/for-organisations/guide-to-data-protection/guide-to-the-general-data-protection-regulation-gdpr/principles/accuracy/> accessed 3 May 2019.
[3] These principles are: (a) lawfulness, fairness and transparency; (b) purpose limitation; (c) data minimisation; (d) accuracy; (e) storage limitation; (f) integrity and confidentiality; (g) accountability (article 5 GDPR)

The accuracy principle has been ever-present in European data protection law. The principle makes a first appearance in early EU Member State data protection instruments in the 1970s, and appears in a list of fair processing principles proposed in 1973 in the US.[5] The principle appeared as a key building block in early international data protection instruments such as the OECD Privacy Principles (1980) and the Council of Europe Convention 108 (1981). Both instruments were recently modernized and retain the accuracy principle.[6] The principle makes a further appearance in the 1995 Data Protection Directive – the first EU level data protection instrument.[7]

The principle has seen few content related changes over the evolution of data protection law – we might call it the crocodile of data protection.[8] The lengthy, uninterrupted, history of the principle underscores the importance of the principle to the data protection edifice. Concretely, the GDPR – in force since early 2016, and applicable since early 2018 – states the accuracy principle, in Article 5(1)(d), as follows: 'Personal data shall be (...) accurate and, where necessary, kept up to date; every reasonable step must be taken to ensure that personal data that are inaccurate, having regard to the purposes for which they are processed, are erased or rectified without delay ("accuracy")'.

In the structure of the GDPR, the accuracy principle plays a triple role. First: the principle outlines the norm that personal data must be: 'accurate and, where necessary, kept up to date'. Second: the principle has an instrumental function in the applicability of other data protection principles. In particular, the principle functions as an applicability criterion for data subject rights to rectify inaccurate personal data – Article 16 – and to object to the processing of inaccurate personal data – Article 18.[9] Third: the principle supports transparency and self-determination principles – outlined in, for example, Article 6(1)(a) concerning the ability to consent and Articles 13 and 14 concerning the right to be informed. Only if data are processed accurately can transparency and self-determination be achieved.[10]

---

[4] Graham Greenleaf, 'Global Data Privacy Laws 2017: 120 National Data Privacy Laws, Including Indonesia and Turkey' [2017] 145 Privacy Laws & Business International Report, 10-13.
<https://ssrn.com/abstract=2993035> accessed 3 May 2019. For example, Article 6 of the 2018 Brazilian law on data protection, contains a list of principles remarkably similar to those in the GDPR. Congresso Nacional Law No. 13,709 Dispõe sobre a proteção de dados pessoais e altera a Lei nº 12.965, de 23 de abril de 2014 (Marco Civil da Internet) 2018, Article 6.
[5] Viktor Mayer-Schönberger, 'Generational Development of Data Protection in Europe' in Phillip Agre and Marc Rotenberg (eds.), *Technology and Privacy: The New Landscape* (MIT Press 1997) 219, 223-224. See, for example, Sweden: Riksdag Datalagen 289 1973. United States Department of Health, Education, and Welfare, *Records, Computers, and the Rights of Citizens* (US Government Printing Office 1973).
<www.justice.gov/opcl/docs/rec- com-rights.pdf> accessed 19 March 2019.
[6] Organisation for Economic Co-operation and Development Guidelines on the Protection of Privacy and Transborder Flows of Personal Data 1980, Art 8; Organisation for Economic Co-operation and Development Guidelines on the Protection of Privacy and Transborder Flows of Personal Data 2013, Article 8; Council of Europe Convention for the Protection of Individuals with regard to Automatic Processing of Personal Data, ETS No. 108, 1981, Art. 5(d); Council of Europe Protocol amending the Convention for the Protection of Individuals with regard to Automatic Processing of Personal Data, CETS No. 223, 2018, Art 7.
[7] Directive 95/46/EC of the Parliament and of the Council of 24 October 1995 on the protection of individuals with regard to the processing of personal data and on the free movement of such data, OJ 1995 L 281/31, Art 6(1)(d).
[8] See also Jiahong Chen, 'The Dangers of Accuracy: Exploring the Other Side of the Data Quality Principle' [2018] 4(1) European Data Protection Law Review 36, 37-38.
[9] This was explicitly highlighted by Advocate General Jääskinen in the CJEU Google Spain case: Case 131/12 *Google Spain SL, Google Inc. v Agencia Española de Protección de Datos and Mario Costeja González* [2014] ECLI:EU:C:2013:424, Opinion of AG Jääskinen, para 104.

The accuracy principle aims to ensure that the individual to whom personal data relate is not subject to misrepresentation, and the consequences of misrepresentation, through their personal data. This aim reflects a jurisprudentially recognised aspect of the right to privacy. As the European Court of Human Rights observed in their *Romet* judgment, the lack of action to correct inaccurate identity information: 'made abuse of...identity by other persons possible, [and] constitutes an "interference" with the...right to respect for..."private life"'.[11] In this regard, the accuracy principle is the only principle in data protection law which directly outlines the normative value of truth in personal data.[12] As a result, it seems likely the principle will only gain in importance in future.

In conclusion, the accuracy principle has been – largely unchanged – a core part of data protection law for at least four decades. But what does it mean for personal data to be held accurately? That question is for the next section.

## III. The Content of the Accuracy Principle

Below, we describe the content of the accuracy principle. We rely on the small hints in the text of the GDPR and the limited relevant case law, sprinkling these with insights from academic work. We distinguish three perspectives on the content of the principle: the scope of the accuracy principle; the concept of the accuracy principle; and the degree of accuracy required by the accuracy principle.

First, what is the *scope* of the accuracy principle? The accuracy principle applies to all personal data processing falling under the scope of the GDPR – eventually, almost all commercial, bureaucratic and scientific processing of personal data. In the words of the Court of Justice of the European Union in the *Google Spain case*: 'all processing of personal data must…comply, first, with the principles relating to data quality set out in Article 6 of Directive 95/46 [Article 5 of the GDPR, including the accuracy principle].'[13] However, the GDPR does grant EU Member States certain power to alter the applicability of the accuracy principle in certain cases. For example, Article 85(2) provides the possibility for Member

---

[10] This observation is especially true given the necessity of holistic accuracy in relation to transparency and self-determination across contexts. How could an individual understand or react to data processing in relation to data doubles which need not correspond to them, or indeed, across contexts, even necessarily to each other?
[11] *Romet v. The Netherlands*, App. No. 7094/06 (ECHR, 14 February 2012), para. 37.
[12] There is an argument that the principle of fairness in Article 5(1)(a) would, even without the accuracy principle in Article 5(1)(d) serve to ensure that, at least in certain cases, personal data were processed accurately and truthfully. This would, however, be a very indirect and weak expression of the normative significance of accuracy in personal data and, given the lack of reference and elaboration of the fairness principle to date in EU data protection law, be likely to be practically overlooked. For a full elaboration of the extent of the fairness principle see: Damian Clifford and Jef Ausloos, 'Data Protection and the Role of Fairness' [2018] Yearbook of European Law 1.
[13] Case C-131/12 *Google Spain SL, Google Inc. v Agencia Española de Protección de Datos and Mario Costeja González* [2014] ECLI:EU:C:2013:424, para 71. This case also references a long history of CJEU case law confirming the point. See, for example: Case C-342/12 *Worten – Equipamentos para o Lar SA v Autoridade para as Condições de Trabalho (ACT)* [2013] ECLI:EU:C:2013:355, para 33.

States to derogate from all principles outlined in Article 5 – including accuracy – for: 'journalistic purposes or the purpose of academic artistic or literary expression'.[14]

Second, what is the *content* of the accuracy principle? Article 5(1)(d) GDPR encompasses two separate concepts of accuracy: factual accuracy and temporal accuracy. Both concepts are intrinsically related, and both can be subsumed by a broad understanding of factual accuracy. Data Protection Authorities state: 'In general, 'accurate' means accurate as to a matter of fact.'[15] In doing so, they foresee the need for personal data to mirror an external 'objective' reality to which they pertain to relate.

Third, what is the *degree* of the accuracy principle? The GDPR's accuracy principle does not require complete accuracy of all personal data in all circumstances. As the Court of Justice of the European Union observed in the *Nowak* case, the purpose of collection and processing will determine the perspective from which accuracy must be judged. They state '[the judgment as to whether] personal data is accurate and complete must be made in the light of the purpose for which that data was collected'.[16] Hence, there must be leeway regarding the degree of accuracy required. If the purpose of processing determines the perspective from which accuracy is to be judged, then the purpose of processing must also determine the degree of accuracy required.

The example of sequenced genomic data for biometric identification illustrates the idea of leeway in the degree of accuracy required by the GDPR. A sequenced genome may contain minor sequencing errors of single data points per million. If such a sequenced genome were used as a biometric identifier, such errors would be insignificant and it would make no sense to consider the data set inaccurate.[17] Controllers must thus take reasonable steps to ensure accuracy in light of the context of processing.[18] Such reasonable steps might be measured in terms of a balance between the potential impacts of processing inaccurate data for data subjects and the financial, organisational and social costs of differing degrees of accuracy for the data controller.[19]

---

[14] See, for a discussion of the breadth of derogations possible under Article 85: Stephan Pötters, 'Artikel 89' in Peter Gola (ed.), *DS-GVO Datenschutz-Grundverordnung VO (EU) 2016/679 Kommentar* (Beck 2017) 781, 785.

[15] Article 29 Working Party, *Guidelines on the Implementation of the Court of Justice of the European Union Judgment on "Google Spain and inc v. Agencia Española de Protección De Datos (AEPD) and Mario Costeja González" C-131/12* (14/EN WP 225, 2014), 15. <http://ec.europa.eu/justice/article-29/documentation/opinion-recommendation/files/2014/wp225_en.pdf> accessed 3 May 2019.

[16] C-434/16 *Peter Nowak v Data Protection Commissioner* [2017] ECLI:EU:C:2017:994, para 53. In this regard, they further observe that it is possible to have personal data which are inaccurate from one perspective nonetheless qualify as being accurate under data protection law. In this case, it was held possible for incorrect – substantively inaccurate – exam answers to nevertheless constitute accurate personal data by virtue of their correct reflection of the quality of the student taking the exam – the purpose for which they were collected. In terms of temporal accuracy the GDPR simply states: personal data must be: 'where necessary, kept up to date'. This principle requires only personal data which, in relation to the purpose of processing, should be kept up to date, to be kept up to date. For example, as the UK Data Protection Authority, the Information Commissioner's Office, observe, it would not be necessary to keep scientific records up to date. See: <https://ico.org.uk/for-organisations/guide-to-the-general-data-protection-regulation-gdpr/principles/accuracy/> accessed 3 May 2019.

[17] See, for example: Dianne Lou, Jeffrey Hussmann, Ross McBee et. al., 'High-throughput DNA sequencing errors are reduced by orders of magnitude using circle sequencing' [2013] 110(49) Proceedings of the National Academy of Sciences USA 19872, 19872.

[18] See also <https://ico.org.uk/for-organisations/guide-to-the-general-data-protection-regulation-gdpr/principles/accuracy/> accessed 3 May 2019.

[19] Whilst the precise factors to be considered and their weighing against each other will, inevitably, be context dependent, there may be some virtue in considering certain philosophical approaches to the identification of

The previous two sections have provided an elaboration of the background and content of the accuracy principle. We now turn to look at the phenomenon in relation to which the accuracy principle will be analysed: opinions. The first step is to provide an overview of the concept of an opinion.

## IV. Outlining the Concept of an Opinion

There is no definition provided for the concept of an opinion – or for synonymous terms such as an inference – in the GDPR, related jurisprudence or even in related areas of law. In literature too, despite the quantity which is currently written about the processing of opinions and inferential personal data – particularly in relation to artificial intelligence and machine learning – there remains surprisingly little effort to define the terms. Addressing this gap, we offer a definition and, on the back of this definition, two subsequent observations – the relevance of these observations will become clear in later sections.

We offer the following definition for the concept of an opinion: 'an assertion about an entity, built on the back of facts about that entity subjected to some interpretative framework to produce new, probable facts.'[20] This definition is a composite definition drawn from definitions for opinions and inferences across a range of disciplines. Inspirations include definitions in natural language understanding, definitions in law – in particular the law of evidence – and definitions in philosophy. There is high correlation between definitions across the diverse sources of inspiration.

An example of this correlation is provided by a comparison of the definitions available in the Oxford English dictionary, that used in the UK law of evidence and that provided by Floridi in his work on the philosophy of information. The Oxford English dictionary defines an opinion as: 'A view or judgement formed about something, not necessarily based on fact or knowledge'.[21] The UK law of evidence defines expert opinion as the application of expert analytical frameworks to the facts of the case to provide supplemental, probable, facts useful for the case at hand.[22] Floridi defines inferential representations as: 'a collection of all that it is known about a category and its members (encyclopaedic knowledge) and that can be used to make predictions or infer non-perceptual information'.[23]

The first observation flowing from our definition is that the concept of an opinion can be broken down into two constituent parts. First, an opinion is built on a set of *available facts* about an entity, known to the opinion holder. This is what serves to differentiate an opinion from a fiction – a statement based on a complete absence of facts would not be an opinion but

---

relevant truth – relevant accuracy – as providing templates for legal calculations of accuracy in data protection law. A particularly relevant approach may be the correctness theory of truth, as been proposed by Floridi. See: Luciano Floridi, *The philosophy of Information* (Oxford 2011) 182-209.

[20] We appreciate the similarity of this construction with those relating to the generation of information from raw data – for example that of Taylor. This should not be a surprise. After all, opinions are information generated from data, simply with a very specific informatic pedigree. Mark Taylor, *Genetic Data and the Law: A Critical Perspective on Privacy Protection* (Cambridge University Press 2012) 42.

[21] See: <https://en.oxforddictionaries.com/definition/opinion> accessed 3 May 2019.

[22] See, for example: *Multiplex Constructions (UK) Limited v Cleveland Bridge UK Limited* [2008] EWHC 2220, para. 672.

[23] Luciano Floridi, *The philosophy of Information* (Oxford 2011) 149.

a fiction. Second, an opinion is built on the interpretation of available facts through a relevant *interpretative framework* to produce probable, albeit unproven, facts.[24]

The second observation flowing from this definition is that, in practice, it may be difficult to distinguish between opinions, facts and falsehoods. Errors may emerge by virtue of epistemic mistakes. For example, an individual may fail to integrate all facts known to them about an entity resulting in a falsehood as opposed to an opinion. Errors may emerge as a result of mistakes in information processing – particularly in complex processing chains. For example, the loss of supplemental data differentiating an opinion from a fact. Finally, errors may emerge as a result of the deliberate obfuscation of a piece of information as an opinion, fact or fiction when the issuing party knows the information to belong in another category.

An example illustrates the difficulty in distinguishing between facts, fictions and opinions. Consider the European Court of Human Rights case of *Khelili v. Switzerland*.[25] In this case, the police had labelled a woman, in official records, as a prostitute. They did so on the back of inferences drawn from business cards found on her person. The label thus reflected an opinion – an inference. Eventually, the opinion turned out to be incorrect. Consider then, if the records had not reflected the genesis of the opinion and the information that the woman was a prostitute appeared as a fact. Other agencies accessing this information would presume factual correctness.

This section has provided an outline of the concept of an opinion. This conceptual outline can now be used as a framework against which to plot the range of uses of the processing of personal data in the form of opinions.

## V. The Broad Range of Uses of Personal Data Opinions

Personal data in the form of opinions about data subjects – personal data opinions – are used in an astounding range of contexts.[26] We distinguish three categories of personal data opinions: (i) human-generated opinions; (ii) algorithm-generated opinions; (iii) human-algorithm-mix generated opinions.

First, all personal data generated solely on the back of human interpretative frameworks will qualify as opinions: human-generated opinions. The category of human-generated opinions includes a broad range of personal data processed for economic and bureaucratic purposes, from employment decisions to social support evaluations to health care evaluations. Human interpretative frameworks may take two different forms: those internal to the human issuing the opinion; and those external to the human issuing the opinion.

Human interpretative frameworks internal to the individual issuing the opinion exist only in that single individual's mind. An example of an opinion generated on the back of an internal interpretative framework would be the opinion considered in the European Court of Human Rights case of *Andreescu v. Romania*. In this case, a politician offered his reasoned opinion – without absolute proof but based on certain known facts – that an opponent had been a collaborator with the communist regime.[27]

---

[24] See also Luciano Floridi, *The philosophy of Information* (Oxford 2011) 132.
[25] *Khelili v. Switzerland*, App. No. 16188/07 (ECHR, 18 October 2011), para. 7-19. See also: Diana Dimitrova, 'Let's get accurate on data accuracy' [2019] (forthcoming).
[26] With this, we do not intend to offer an exhaustive list of all types of opinion personal data. Rather, we intend to offer an overview, and with it a taste of the breadth, of processing of personal data which constitute opinions.

Human interpretative frameworks external to the human issuing the opinion are standardised, proceduralised thought processes for the interpretation of facts in a context. An example of an opinion generated on the back of an external interpretative framework would be the opinion in the Court of Justice of the European Union case *YS. and M. and S* concerning the legal residency status of the plaintiff. This opinion was the result of an interpretation of facts through a legal framework – an interpretative framework created and imposed by the state.[28] As Wachter and Mittelstadt observe: 'A legal analysis is comparable to [an] analysis…from which new data is derived or inferred.'[29]

Second, almost all personal data generated solely on the back of algorithmic interpretative frameworks will qualify as opinions: algorithm-generated opinions.[30] The category of algorithm-generated opinions will cover a broad range of personal data processed, predominantly, for economic purposes. The category is particularly relevant in relation to online profiling activities such as behavioural advertising.[31]

Algorithmic inferences emerge on the back of factual – or supposedly factual – datasets subjected to a programmed interpretative framework with the intention of producing probabilistic conclusions about the individuals represented by those data sets. In this regard, an algorithmic interpretative framework may produce opinions applicable to individuals who are not – and indeed may never need have been – represented within the originally collected and analysed dataset.

Algorithmic interpretative frameworks may be programmed with the intention of producing opinions of specific informational content. Algorithmic interpretative frameworks may also be built around artificial intelligence and machine learning principles to recognise general patterns in data sets. Indeed, the aim of machine learning is to find hidden relations or patterns in data sets.[32]

Third, all personal data generated on the back of human-algorithm-mix interpretative frameworks will qualify as opinions. This will be the case regardless of which interpretative framework – whether human or algorithmic – eventually produces the final opinion. At this stage in the development of automated decisions and algorithmic frameworks, however, the human decision-maker will usually be the final point of analytical instance. This category of opinions will cover a broad range of personal data generated for both economic and bureaucratic purposes.

---

[27] *Andreescu v. Romania*, App no 19452/02 (ECHR, 8 June 2010).
[28] Joined Cases C-141/12 and C-372/12 *YS v Minister voor Immigratie, Integratie en Asiel and Minister voor Immigratie, Integratie en Asiel v M, S* [2014] ECLI:EU:C:2014:2081.
[29] Sandra Wachter and Brent Mittelstadt, 'A Right to Reasonable Inferences: Re-Thinking Data Protection Law in the Age of Inferences and Big Data' [2019] *Columbia Business Law Review* 1, 21 (forthcoming).
[30] The only exception will be algorithm generated personal data which, either by design or chance, produce logically certain results.
[31] See generally: F.J. Zuiderveen Borgesius, *Improving privacy protection in the area of behavioural targeting* (Kluwer Law International 2015).
[32] Lehr and Ohm describe machine learning as follows: "machine learning refers to an automated process of discovering correlations (sometimes alternatively referred to as relationships or patterns) between variables in a dataset, often to make predictions or estimates of some outcome." David Lehr and Paul Ohm, 'Playing with the data: What legal scholars should learn about machine learning' [2017] 51 UCDL Rev 653, 671. See also Raphael Gellert, 'Comparing the definitions of data and information in data protection law and machine learning: a useful way forward for algorithmic regulation?' [2018] SSRN Working Paper 1, 15. <https://papers.ssrn.com/sol3/Delivery.cfm/SSRN_ID3284493_code3218171.pdf?abstractid=3284493&mirid=1> accessed 3 May 2019.

An example of a process involving human-algorithm-mix opinions is the system imagined in the iBorderCtrl project.[33] In the project, algorithms are envisaged as being responsible for the initial screening of potential immigrants into the EU – via the algorithmic analysis of veracity based on an interview – prior to travel. In the case of suspicion – highlighted by the algorithmic interview interpretation – at the EU border, human border guards will conduct supplemental interviews and will take a final decision about the permissibility of a traveller's entry into the EU.

Against this background, the next section addresses the first part of the original research question: are opinions subject to data protection law's accuracy principle?

## VI. Opinions are the Subject of the Accuracy Obligation

We argue that the accuracy principle applies to opinions. In demonstrating our argument, we consider the applicability of the principle to the two building blocks of an opinion: (i) facts; (ii) interpretative frameworks.

Asserting that facts should be considered in terms of the accuracy principle is uncontroversial. As discussed above, the accuracy principle applies to the processing of all facts. The processing of facts which form the basis of an opinion is no exception. This logic is unaffected by the subsequent addition of an interpretative framework to these facts to create an opinion. As the UK Data Protection Authority states in relation to the accuracy principle and credit scores, for example: '[the data controller] must ensure the accuracy (and adequacy) of the underlying data'.[34]

Asserting the accuracy principle is applicable to opinions generated on the back of interpretative frameworks, however, is more complex. A significant hurdle appears. The most commonly voiced objection, for example by scholars such as Herbst, to the idea that the accuracy principle can apply to opinions runs as follows: as opinions rely on interpretative frameworks, this means they do not relate to an objective reality and therefore lie beyond the scope of accuracy and inaccuracy calculations. By extension, this objection asserts that opinions cannot be considered as inaccurate.[35] This objection is superficially reasonable. A deeper look at the objection, however, shows it to be flawed.

The objection relies on the blunt assumption that there is a clear line between facts, which can be defined in terms of accuracy and inaccuracy and opinions, based on interpretative frameworks, which cannot. This is not true. It is true that no interpretative framework can produce opinions offering epistemic certainty. It is also true, however, that certain interpretative frameworks, in certain contexts, are epistemologically more trusted to provide reliably accurate information than others. Interpretative frameworks can be considered in terms of degrees of accuracy. Thus, in terms of personal data opinions, certain interpretative frameworks will be trusted to provide more reliably accurate personal data than others.

---

[33] Algorithm Watch, *Automating Society Taking Stock of Automated Decision-Making in the EU* (Report, 2018) 37. <https://algorithmwatch.org/wp-content/uploads/2019/01/Automating_Society_Report_2019.pdf> accessed 3 May 2019.
[34] <https://ico.org.uk/for-organisations/guide-to-data-protection/guide-to-the-general-data-protection-regulation-gdpr/principles/accuracy/> accessed 3 May 2019.
[35] See, for example: Tobias Herbst, 'Art. 5 Grundsätze für die Verarbeitung personenbezogener Daten' in Jürgen Kühling and Benedikt Buchner (eds*.), DatenschutzGrundverordnung/ BDSG* (2nd Edition Beck 2018) 229, para 60; Sebastian Dienst, 'Lawful processing of personal data in companies under the GDPR' in Daniel Rücker and Tobias Kugler (eds.), *New European General Data Protection Regulation: A Practitioner's Guide* (Beck/Hart/Nomos 2018) 68, para 326.

For example, in the clinical context, the diagnostic opinion of a trained doctor – using an interpretative framework based on years of medical school and experience – will generally be trusted to be more accurate than that of a layperson – using an interpretative framework based on common knowledge. This differentiation is based on assumptions as to the reliable accuracy of information produced by medical interpretative frameworks. This differentiation is so normatively accepted as to be the justification for legal penalties – for example, in the UK Medical Act 1983 for the impersonation of physicians.[36]

We have already established, above, that the GDPR does not insist on absolute accuracy in personal data. In this regard, one can thus look at the interpretative frameworks behind personal data opinions in terms of whether they furnish information which is 'accurate enough' in a given context. Many areas of law already draw such a line. Returning to the medical context, one example would be the law of medical negligence and the clinician's duty to rely on professionally accepted interpretative frameworks in providing diagnoses.[37] For example, the use of smell as an interpretative framework for the diagnosis of broken ribs would be scientifically and medically unsound. Information produced on the back of this framework would not reach the required reliability threshold to be 'accurate enough'.

The GDPR is omnibus legislation. Accordingly, in contexts in which accuracy rules concerning interpretative frameworks around personal data opinions already exist, these will inform the normative standards of 'accurate enough' under the GDPR.[38] In relation to medical opinions, therefore, prevailing standards of requisite accuracy of interpretative frameworks in that area of law will provide a guideline for standards under the GDPR. In situations where no standards exist – as with many principles in the GDPR in many processing contexts – factors to be considered in drawing the line of 'accurate enough' in relation to interpretative frameworks will be context dependent.[39]

In sum, opinions can be the subject of the accuracy obligation. We now move on to the second part of the research question: *should* opinions be subject to data protection law's accuracy principle?

# VII. Opinions should be the Subject of the Accuracy Principle

---

[36] See, for example: Medical Act 1983 (UK), s 49(1);
[37] See, for a discussion of the law of medical negligence and the accuracy standards imposed on the use of interpretative frameworks in clinical care: Daniele Bryden and Ian Storey, 'Duty of care and medical negligence' [2011] 11(4) British Journal of Anaesthesia 124, 125.
[38] Dara Hallinan, Feeding Biobanks with Genetic Data: What role can the General Data Protection Regulation play in the protection of genetic privacy in research biobanking in the European Union? (PhD Thesis Vrije Universiteit Brussel 2018) 324-325.
[39] See, for an extended discussion of the requirement to engage in context specific considerations of action in relation to multiple provisions of the GDPR: Damian Clifford and Jef Ausloos, 'Data Protection and the Role of Fairness' [2018] Yearbook of European Law 1, 25-48.

We argue the accuracy principle should apply to personal data opinions. We put forward three arguments supporting our position. The first argument highlights the applicability of the fundamental logic behind the accuracy principle to personal data opinions. The second and third arguments refute two objections which might be put forward as reasons for not applying the accuracy principle to personal data opinions.

As discussed above, the accuracy principle aims to ensure individuals are not falsely represented via their data doubles. Accordingly, the principle ensures individuals are not irrationally or unfairly judged based on false representations. The logic behind the principle undoubtedly applies to the processing of personal data in the form of opinions. Opinions are regularly developed and deployed to represent and to judge individuals. As discussed, bureaucratic opinions on individuals are prevalent in society.

Indeed, in some of the contexts discussed above, opinions will be the basis of decisions with highly significant life consequences. For example, credit reporting typically relies on profiling individuals to estimate their creditworthiness. Creditworthiness opinions can have effects on individuals' lives in terms of whether they can secure credit – for example for a bank loan for the purchase of a house. Creditworthiness opinions may also have effects on individuals' lives as credit scores are used as a proxy to judge future financial reliability – for example as proof of reliability for a landlord when renting a flat. Hence, the rationale for the accuracy principle also applies to personal data in the form of opinions.

A first argument against the applicability of the accuracy principle to personal data opinions might be put forward on the grounds the principle is simply superfluous. Two versions of this argument might be elaborated. The first version holds the principle to be superfluous on economic grounds: certain controllers' bottom line may be negatively affected by processing inaccurate opinions so they will strive for accuracy anyway – for example in the insurance context.[40] The second version of the argument holds the principle to be superfluous on legal grounds: certain controllers are subject to other legal obligations requiring accuracy – for example under biomedical research law.[41]

This argument does not stand up to scrutiny. First, neither of the versions of the argument presents a fundamental challenge to the applicability of the accuracy principle in relation to opinions: superfluous obligations are not problematic obligations. In turn, even if the arguments did present a fundamental challenge to the applicability of the obligation, they are contextually specific and fail to constitute a challenge to the general relevance of the principle to all personal data opinions.

A second argument against the applicability of the accuracy principle to personal data opinions might be put forward on the grounds that the principle could have disproportionate impacts on other rights and legitimate interests. One obvious example would be the right to freedom of expression in relation to political speech.

Indeed, the ability to make unproven – potentially mistaken – claims in relation to political speech has been expressly recognised in European Court of Human Rights case law even in relation to particularly harsh claims. The case of *Kusbaszewski v. Poland* provides an example.[42] In this case, a local councillor in the Kleczew municipality in Poland publicly expressed his opinion that fellow councillors had engaged in money laundering. The Court recognised the harshness of the allegations. The Court ultimately, however, recognised the necessity of the ability for politicians to make such allegations in the interests of public debate.

---

[40] See, for example: Ruth Stirton, 'Insurance, Genetic Information and the Future of Industry Self-Regulation in the UK' [2012] 4(2) Law, Innovation and Technology 212.
[41] See, for example, Article 22(5) of the Estonian Human Genes Research Act: Human Genes Research Act 2000 (Estonia), Art. 22(5). Unofficial English translation available at <https://www.riigiteataja.ee/en/eli/531102013003/consolide> accessed 3 May 2019.

In such cases, the imposition of a strict accuracy principle would be problematic. As noted above, however, the degree of accuracy required by the GDPR's accuracy principle is context dependent. Any consideration of relevant accuracy must thus consider the apposite rights and interests of the parties involved. Thus, if opinions with a certain margin of error are appropriate in context, such opinions need not fall foul of the accuracy principle. Eventually, as Chen notes, the degree of accuracy necessary in any context must always be the subject of ethical reflection.[43]

In conclusion, the accuracy principle is important because the principle aims to protect people against decisions based on wrong information. That rationale also applies to opinions. Arguments disputing the applicability of the principle are unconvincing.

# VIII. Conclusion

The overwhelming majority of information privacy and data protection statutes and treaties around the world include a version of an accuracy principle. Yet, for some types of personal data, the applicability of the principle is not clear. One such case is that of personal data in the form of opinions about data subjects.

This paper thus focused on the following question: are opinions subject to data protection law's accuracy principle, and should they be? We analysed both pro and contra arguments. Ultimately, we concluded that data protection law's accuracy principle does, and should, apply to personal data consisting of opinions about individuals.

In terms of the applicability of the accuracy principle to personal data opinions: the logic of applicability can be shown by considering how both building blocks of an opinion – facts and interpretative frameworks – can be considered in terms of the principle. Regarding facts, the situation is uncontroversial. All facts can and must be treated accurately in data protection law. Regarding interpretative frameworks, certain interpretative frameworks produce more accurate opinions than others. One can – indeed this is legally common – in a given context, draw lines between interpretative frameworks capable of producing sufficiently accurate opinions and those incapable of producing sufficiently accurate opinions.

In terms of the normative legitimacy of the applicability of the principle to opinions on individuals: the fundamental rationale behind the accuracy principle – to protect individuals from being misrepresented through data as well as the consequences of this misrepresentation – is applicable in relation to personal data opinions. There are no practical or normative arguments which call this assertion into doubt. Nor are there any practical or normative arguments highlighting countervailing interests which should override the need for accuracy.

---

[42] *Kusbaszewski v. Poland* App no 571/04 (ECHR, 2 February 2010) par. 38-43.
[43] Jiahong Chen, 'The Dangers of Accuracy: Exploring the Other Side of the Data Quality Principle' [2018] 4(1) European Data Protection Law Review 36, 40-50.